\documentstyle[psfig]{caosp}

\def\Fe2{\hbox{{\rm Fe}~{$\scriptstyle {\rm II}$}}}
\def\Cr2{\hbox{{\rm Cr}~{$\scriptstyle {\rm II}$}}}

\def\degr{^{\circ}}
\begin{document}

\title{Multi-Element Doppler Imaging of the Ap star~~~ $\epsilon$ Ursae Majoris}
\author{T. L\"uftinger \and R. Kuschnig \and{W.W.Weiss}}
\institute{Institute for Astronomy, University of Vienna, Austria}

\maketitle

\begin{abstract} 
The surface distribution of five elements: $iron$, $chromium$, $titanium$, 
$magnesium$ and $manganese$ on the magnetic A0pCr star $\epsilon$ UMa, have
been calculated using the Doppler imaging technique. 
We found that $iron$, $chromium$ and $manganese$ are correlated with the 
assumed dipole magnetic field geometry of this star,
which is apparently not the case of $magnesium$ and $titanium$. 

\end{abstract}


\section{Introduction}

The scientific goal of applying the Doppler imaging technique to Ap
stars, is to provide observational constraints on the diffusion mechanism
in the presence of a global magnetic field. $\epsilon$ UMa (HD 112185, HR
4905), an A0pCr star, is known as the brightest member (V=1.77) of the
class of peculiar A type stars. Bohlender and Landstreet (1990) measured
a weak, reversing magnetic field for $\epsilon$ UMa, that appears to be
dominated by a dipole component with a polar magnetic field strength in
the order of 400 Gauss.  Furthermore, maps of $chromium$, $iron$ (Rice \&
Wehlau, 1997), $oxygen$ and $calcium$ (Babel et al., 1995) have been
published, whereby the distribution of each of these elements appears to
be correlated with the assumed dipole magnetic field geometry. 

\section {Observations}

Observations of $\epsilon$ UMa were done in June 1994 and in March 1995 at
the Observatoire de Haute-Provence using the spectrograph AUR\'ELIE (attached
to the 1.52-m telescope) in two spectral regions: 4060 - 4260 \AA \
  and 4440 - 4640 \AA. The spectral resolution is about 20000 and the
Signal-to-Noise ratio above 150. 

\section{Input data for Doppler Imaging}

The maps were calculated by using the surface imaging technique described
by Piskunov and Rice (1993). A Kurucz ATLAS9 model atmosphere with
$T_{\rm eff} = 9500$~K and $\log g=3.6$ was used for computing the local
line profile tables. The value of $v\sin i$ was assumed to be $35$~km\,s$^{-1}$
and the inclination angle $i$ was chosen as 45 $\degr$ . Both values give
a minimal deviation of the observed line profiles from the computed
ones. The rotation phases of the spectra we used for mapping were
computed according to: $JD = 2434131.124 + 5^{d}.0887\,E$.

\section{Discussion}

The surface abundance distributions of the five elements we treated can
be divided into two groups. 

The $iron$ and $chromium$ (Figure 1)
distributions show a clear depleted band which coincides with the assumed 
magnetic equator, confirming the results of Rice \& Wehlau.  The $manganese$ distribution is very similar to that of these elements,
which accumulate near the magnetic poles. They are all slightly
overabundant compared to solar values: $manganese$ and $chromium$ are
about 0.8 dex above solar values, while $iron$ is 1 dex above.  

\begin{figure} 
\centerline{\psfig{figure=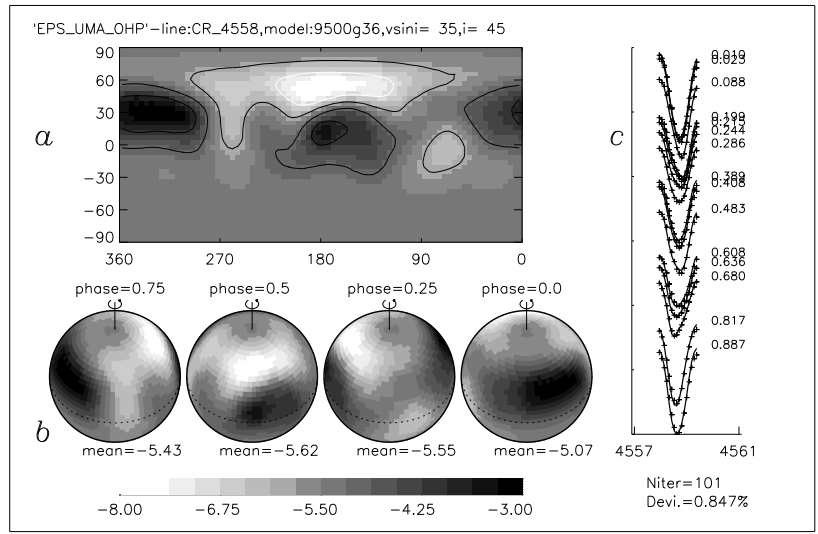,height=7cm}}
\caption{The chromium abundance distribution of $\epsilon$ UMa was
obtained from the Cr\,{\sc ii}, 4558 \AA\ line. This element appears to be on
average, about 0.8 dex more abundant than in the Sun.} 
\end{figure}

However, the
$titanium$ (Figure 2) and $magnesium$ surface structures have much less
contrast in terms of peak-to-peak abundances and are apparently not
significantly correlated with the magnetic dipole geometry. So far, seven
different elements have been mapped for $\epsilon$ UMa. 

Together with the
$oxygen$ and $calcium$ maps published by Babel et al. (1995), which reveal
abundance enhancements located at the magnetic equator, the maps of
$iron$, $chromium$, $magnesium$, $manganese$ and $titanium$ provide important
constraints for building models of diffusion in the presence of a global
magnetic field. This should provide a better understanding of the
hydrodynamics in the atmospheres of Ap stars. 

\begin{figure} 
\centerline{\psfig{figure=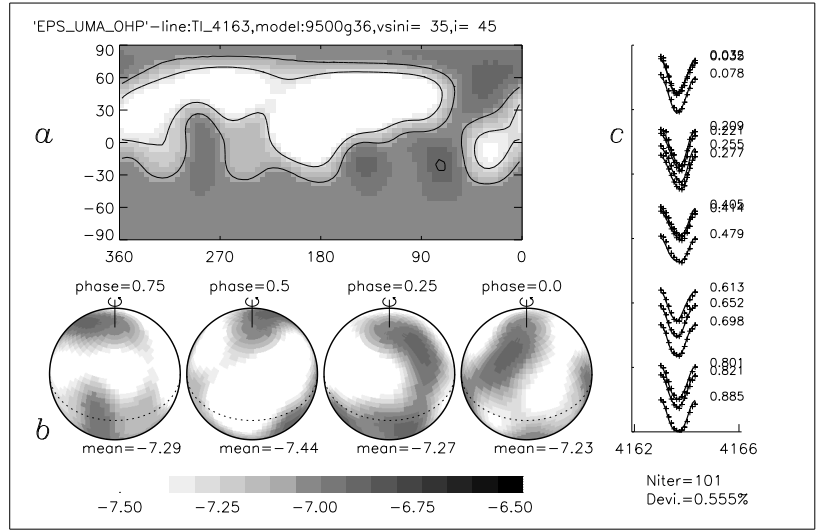,height=7cm}}
\caption{Titanium abundance distribution of $\epsilon$ UMa. The Ti\,{\sc ii},
4163 \AA\ line was used for the inversion procedure. Titanium is on
average slightly depleted on the surface of this star compared to solar
abundance.} 
\end{figure}

\end{document}